\documentclass[10pt,letterpaper]{article}

\usepackage[utf8]{inputenc}
\usepackage[T1]{fontenc}

\usepackage{amsmath,amsfonts,amssymb,amsthm}

\usepackage{graphicx}
\usepackage{booktabs}
\usepackage{array}

\usepackage[linesnumbered,ruled,vlined]{algorithm2e}

\usepackage[colorlinks=true,linkcolor=blue,citecolor=blue,urlcolor=blue]{hyperref}

\usepackage[letterpaper,margin=1in]{geometry}

\usepackage{microtype}

\usepackage{enumitem}

\theoremstyle{definition}

\title{\LARGE \textbf{Modeling GRNs with a Probabilistic Categorical Framework}}

\author{
Yiyang Jia\textsuperscript{1,*}, 
Zheng Wei\textsuperscript{1}, 
Zheng Yang\textsuperscript{2}, 
Guohong Peng\textsuperscript{3}
\\[0.5em]
\textsuperscript{1}Tokyo City University\\
\textsuperscript{2}Sichuan University\\
\textsuperscript{3}XiChang University\\[0.5em]
\textsuperscript{*}Correspondence: kaiyou@tcu.ac.jp
}

\date{}

\begin{document}

\maketitle

\begin{abstract}
Understanding the complex and stochastic nature of Gene Regulatory Networks (GRNs) remains a central challenge in systems biology. Existing modeling paradigms often struggle to effectively capture the intricate, multi-factor regulatory logic and to rigorously manage the dual uncertainties of network structure and kinetic parameters. In response, this work introduces the Probabilistic Categorical GRN (PC-GRN) framework. It is a novel theoretical approach founded on the synergistic integration of three core methodologies. Firstly, \textbf{category theory} provides a formal language for the modularity and composition of regulatory pathways. Secondly, \textbf{Bayesian Typed Petri Nets (BTPNs)} serve as an interpretable, mechanistic substrate for modeling stochastic cellular processes, with kinetic parameters themselves represented as probability distributions. The central innovation of PC-GRN is its end-to-end generative Bayesian inference engine, which learns a full posterior distribution over BTPN models ($P(G, \Theta | D)$) directly from data. This is achieved by the novel interplay of a \textbf{GFlowNet}, which learns a policy to sample network topologies, and a \textbf{HyperNetwork}, which performs amortized inference to predict their corresponding parameter distributions. The resulting framework provides a mathematically rigorous, biologically interpretable, and uncertainty-aware representation of GRNs, advancing predictive modeling and systems-level analysis.
\end{abstract}


\vspace{1em}

\section{Introduction}
\label{sec:introduction}

Gene Regulatory Networks (GRNs) represent the complex control circuitry that governs the response of living cells to internal and external cues~\cite{Gene_regulatory_networks}. These networks, composed of genes, their products, and the regulatory interactions between them, are characterized by non-linear dynamics, feedback loops, and a modular architecture~\cite{Current_approaches_to_GRNs,modeling_and_analysis_of_GRNs}. Understanding their structure and function is therefore a central goal in systems biology. While seminal theoretical work has established key principles of network motifs and design~\cite{alon2019introduction,tyson2019modeling}, inferring these complex, dynamic systems from noisy experimental data remains a challenging scientific challenge.

\subsection{The Challenge: Integrating Structure, Dynamics, and Uncertainty}
Modeling GRNs effectively requires navigating a landscape of inherent complexities~\cite{modeling_and_analysis_of_GRNs}. These challenges include a vast combinatorial space of potential interactions (\textbf{high dimensionality})~\cite{Huang2005}, the synergistic effects of multiple regulators (\textbf{non-linearity})~\cite{Simon2007}, and the pervasive influence of random fluctuations (\textbf{stochasticity})~\cite{Grant2006}. Consequently, the goal of modern GRN inference is not to identify a single "correct" network, but rather to characterize a \textbf{posterior distribution} over an ensemble of plausible models.

Existing modeling paradigms, however, face difficult trade-offs. \textbf{Boolean Networks} offer scalability at the cost of quantitative detail; \textbf{Differential Equation} models provide mechanistic realism but struggle with parameterization for large systems~\cite{DE}; and traditional \textbf{Bayesian Networks} handle uncertainty well but are constrained to acyclic graphs, precluding essential feedback loops~\cite{Bayesian_Networks}. While formal frameworks for compositionality exist~\cite{aduddell2024compositional}, a unified, end-to-end learning approach that can infer these complex, cyclic, and probabilistic structures directly from data, remains missing as a critical gap. 

\subsection{Our Contribution: The PC-GRN Framework}
This paper introduces the \textbf{Probabilistic Categorical Gene Regulatory Network (PC-GRN)} framework to address this gap. PC-GRN is founded on the synergistic integration of three powerful formalisms from mathematics and computer science:

\begin{itemize}
    \item \textbf{Category Theory}, which provides a formal and abstract language for describing modularity, hierarchy, and the rules of composition within complex systems.
    
    \item \textbf{Bayesian Typed Petri Nets (BTPNs)}, which we employ as interpretable, mechanistic models for the stochastic and concurrent dynamics of molecular processes. Kinetic parameters in BTPNs are themselves probability distributions, inherently capturing parametric uncertainty.
    
    \item An \textbf{End-to-End Bayesian Learning Architecture}, which employs generative models (GFlowNets) and dynamic parameterization (HyperNetworks) to infer a posterior distribution over both the structure and parameters of BTPN models from data.
\end{itemize}

The novelty of PC-GRN lies not only in the application of these tools, but in their deep integration, as summarized in Table~\ref{tab:methodologies}. Category theory provides the formal structure for composing the probabilistic, mechanistic models instantiated by BTPNs. This compositional structure, in turn, makes the learning problem scalable via the decompositional approach used by our generative Bayesian engine. The result is a unified framework where each component enhances the capabilities of the others, yielding insights beyond the reach of any single approach.

\begin{table}[h!]
\centering
\caption{Core Methodologies and Their Contributions to the PC-GRN Framework.}
\label{tab:methodologies}
\begin{tabular}{|>{\raggedright\arraybackslash}p{0.25\textwidth}|>{\raggedright\arraybackslash}p{0.7\textwidth}|}
\hline
\textbf{Methodology} & \textbf{Description and Contribution to PC-GRN} \\
\hline
Category Theory & Provides the formal algebra of composition. In PC-GRN, it is used to define a path category over an influence graph, where morphisms represent regulatory pathways and composition is governed by formal rules ($\circ_T, \circ_P$) that propagate type and uncertainty. 
\\
\hline
Generative Bayesian Inference & Provides the engine for learning from data and quantifying uncertainty. This is realized via a synergistic architecture where a GFlowNet samples network topologies ($G$) and a HyperNetwork predicts their parameter distributions ($\Theta$), together learning the joint posterior $P(G, \Theta | D)$. \\
\hline
BTPNs & Serve as the interpretable and simulatable mechanistic model. A BTPN describes the stochastic dynamics of molecular processes, and its kinetic parameters are themselves probability distributions, explicitly representing epistemic uncertainty. It provides the physics engine for computing the reward signal. \\
\hline
\end{tabular}
\end{table}

This paper provides a comprehensive theoretical foundation for the PC-GRN framework. We begin by providing the necessary theoretical background (Section 2) before formally defining the three layers of the PC-GRN formalism (Section 3). We then detail the end-to-end Bayesian learning architecture and its training algorithm (Section 4), and conclude with a discussion of the framework's implications, limitations, and directions for future research (Section 5 and 6).
\section{Theoretical Background and Related Research}
\label{sec:background}

This section first presents a comprehensive review of prevailing GRN modeling paradigms, critically assessing their respective strengths and limitations. Against this backdrop, we then introduce the three foundational pillars of the PC-GRN framework. Our approach is built on the synergistic integration of category theory for compositional algebra, generative Bayesian inference for learning under uncertainty, and BTPNs for interpretable mechanistic modeling. This synthesis of formalisms is designed to cohesively address the modular architecture, stochastic dynamics, and compositional nature of GRNs, providing a rigorous foundation for the novel learning architecture detailed in subsequent chapters.

\subsection{An Overview of GRN Modeling Paradigms}
\label{ssec:overview_paradigms}

GRN modeling techniques are typically distinguished by how they represent system states(discrete or continuous) and how they handle dynamics—deterministic or stochastic. Each perspective offers unique insights into the complexities of biological networks. However, no existing paradigm fully addresses the compositionality, hierarchical structure, and uncertainty that characterize gene regulation. These models exhibit limitations when addressing the multifaceted nature inherent in complex biological systems. For a comprehensive survey of GRN modeling paradigms, including single-molecule and hybrid models, we refer the reader to Figure~1 and Table~1 in \cite{vijesh2013modeling}.

\subsubsection{Discrete and Logical Models: Boolean Networks}

Boolean Networks (BNs) represent one of the simplest modeling approaches, reducing gene states to binary values: ``on'' or ``off''~\cite{xiao2009tutorial}. The state of each gene evolves over discrete time steps according to logical functions applied to its regulatory inputs~\cite{lahdesmaki2003learning}. This abstraction provides clarity and computational efficiency, making BNs particularly suitable for qualitative analyses such as identifying attractors or stable states within the system~\cite{albert2004boolean,de2006qualitative}. However, their binary nature limits the representation of graded, dose-dependent gene expression commonly observed in biological systems~\cite{puvsnik2022review}. While probabilistic extensions of BNs have been proposed~\cite{shmulevich2002boolean}, the basic deterministic update rules do not adequately capture the intrinsic stochasticity of molecular interactions, including noise and low copy-number effects~\cite{puvsnik2022review}. Moreover, the typical assumption of synchronous updates across all genes introduces a further idealization, which may deviate from the inherently asynchronous and time-delayed behavior of real cellular processes~\cite{bornholdt2008boolean,marku2023time}.

\subsubsection{Continuous and Mechanistic Models: Differential Equations}

Differential equation (DE) models, especially those employing ordinary differential equations (ODEs), describe the continuous evolution of molecular concentrations by modeling their rates of change over time~\cite{parmar2017time}. Grounded in chemical kinetics (e.g., mass-action and Michaelis-Menten laws), these models faithfully capture the nonlinear, continuous dynamics of biochemical systems~\cite{yang2020overview}. As a result, they often yield more realistic representations of cellular behavior~\cite{karlebach2008modelling}. Nonetheless, DE models face notable challenges. Their reliance on numerous kinetic parameters(such as reaction rates and binding affinities) makes parameter estimation difficult, especially in the presence of noisy and sparse experimental data~\cite{yang2020overview}. Additionally, the computational burden of simulating large, intricate networks often restricts the application of these models to relatively small and well-characterized systems~\cite{yang2020overview}.

\subsubsection{Probabilistic Graphical Models: Bayesian Networks}

Bayesian Networks (BNs) represent conditional dependencies among genes through directed acyclic graphs, with edges denoting probabilistic influence~\cite{liu2016inference}. Their probabilistic foundation enables robust modeling under uncertainty and facilitates the incorporation of prior biological knowledge. However, two key limitations constrain their applicability to GRNs. First, the requirement of acyclic structure precludes feedback loops and cyclic interactions, which are essential features of biological regulation~\cite{banf2017computational}. Second, conventional BNs lack the ability to capture temporal dynamics, rendering them unsuitable for modeling changes in gene expression over time~\cite{yu2017inference}. Dynamic Bayesian Networks (DBNs) address this shortcoming by unrolling dependencies across discrete time points, thereby enabling the representation of temporal processes and cyclic dependencies~\cite{zou2005new, dondelinger2012dynamic}. Yet, this additional expressive capacity incurs significant computational cost, often limiting DBNs to small-scale systems~\cite{zou2005new}.

\subsubsection{Connectionist and Data-Driven Models: Neural Networks}

Neural networks (NNs), particularly recurrent neural networks (RNNs), emerged as powerful data-driven models for GRN inference~\cite{xu2007inference,raza2016recurrent,shu2021modeling}. In these approaches, gene expression levels serve as inputs, while learned weights reflect regulatory interactions. The recurrent structure is well-suited to capturing temporal dependencies and feedback loops within time-series data~\cite{shu2021modeling}. NNs exhibit impressive flexibility, capable of approximating complex nonlinear relationships and demonstrating robustness to noisy inputs~\cite{shu2021modeling, zhao2022hybrid}. Nevertheless, they come with substantial challenges. Model performance is highly sensitive to architecture and hyperparameter choices; training can be data-intensive and susceptible to local minima~\cite{raza2016recurrent}. Furthermore, the ``black-box'' nature of many neural models hinders biological interpretability, as internal parameters often lack transparent mechanistic meaning.

Taken together, these paradigms entail fundamental trade-offs~\cite{vijesh2013modeling,zhao2021comprehensive}:
\begin{itemize}
    \item Boolean models prioritize simplicity and scalability but forgo quantitative detail.
    \item Differential equation models offer mechanistic fidelity but suffer from parameter complexity and limited scalability.
    \item Bayesian models provide probabilistic robustness while struggling with feedback modeling or computational scalability.
    \item Neural networks enable expressive, nonlinear modeling at the expense of interpretability and training efficiency.
\end{itemize}

Faced with these trade-offs, a natural question arises: can a framework be developed that integrates the complementary advantages of these methods while mitigating their individual limitations? The PC-GRN framework is proposed as a step in this direction.

\subsection{Conceptual Foundations of the PC-GRN Framework}
\label{ssec:core_components}

Our PC-GRN framework is founded on the synergistic integration of three fundamental formalisms. Each contributes unique and complementary strengths, collectively forming a unified framework that bridges high-level compositional algebra with data-driven mechanistic modeling.

\subsubsection{Category Theory: The Algebra of Composition}
Category theory provides the abstract mathematical language for describing structure and composition. In PC-GRN, it is not only a representational tool but the formal algebra that governs how network components interact. We use it to:
\begin{itemize}
    \item Define GRNs as a \textbf{probabilistically enriched path category}, where objects are genes and morphisms are regulatory pathways.
    \item Formalize the \textbf{composition rules} ($\circ_T, \circ_P$) that describe how the type and uncertainty of a multi-step regulatory cascade are calculated from its individual steps.
\end{itemize}
This provides a principled and scalable foundation for reasoning about the hierarchical and modular nature of GRNs.

\subsubsection{Generative Bayesian Inference: The Engine for Learning from Data}
To learn the network from data, PC-GRN employs a sophisticated, end-to-end Bayesian inference engine. This goes beyond traditional Bayesian methods by learning a \textbf{generative model for the posterior distribution}, $P(G, \Theta | D)$. This is realized through the interplay of:
\begin{itemize}
    \item A \textbf{GFlowNet}, which learns a policy to sample network topologies ($G$) from the vast combinatorial space of possibilities.
    \item A \textbf{HyperNetwork}, which performs amortized inference to predict the kinetic parameter distributions ($\Theta$) for any given topology.
\end{itemize}
This approach allows the framework to integrally handle both structural and parametric uncertainty, learning an entire ensemble of plausible models rather than a single point estimate.

\subsubsection{BTPNs: The Interpretable Mechanistic Model}
To ground the framework in biological reality, we use BTPNs as the mechanistic substrate. A BTPN models the stochastic and concurrent dynamics of molecular processes, where:
\begin{itemize}
    \item \textbf{Places} and \textbf{Transitions} have clear biological correlates (genes/molecules and reactions), ensuring model interpretability.
    \item The model is inherently \textbf{Bayesian}, as its kinetic parameters are not fixed values but are themselves probability distributions ($\Theta$). This allows the mechanistic model to explicitly represent our uncertainty about the precise rates and strengths of interactions.
\end{itemize}
The BTPN thus serves as the executable, mechanistic model for evaluation. Its stochastic simulation, using the structure and parameters proposed by the GFlowNet and HyperNetwork, generates the data required for computing the reward signal that drives the entire learning process.
\section{The PC-GRN Framework: Formalism and Learning Architecture}
\label{sec:framework}
This section presents the complete technical foundation of the PC-GRN framework, illustrated in Figure~\ref{fig:framework}. As the diagram shows, the framework's design is partitioned into three conceptual parts. The Three-Layer Formalism (top panel) defines the mathematical objects of our framework, establishing a clear progression from a mechanistic BTPN as the uncertainty-aware model, through an abstracted influence graph, to a high-level compositional category ($G_{\mathrm{prob}}$). This formalism is instantiated from data by the End-to-End Bayesian Learning engine (middle panel), a synergistic system of generative models and mechanistic simulators designed to learn the full posterior distribution over BTPN models, capturing uncertainty in both structure ($G$) and parameter distributions ($\Theta$). The final output is the characterization of this learned Posterior Distribution (bottom panel), which enables a rich, multi-faceted quantification of the discovered network's structural and parametric uncertainty. The following sections will formally define each of these components in detail.

\begin{figure}[h!]
    \centering
    \includegraphics[width=0.8\textwidth]{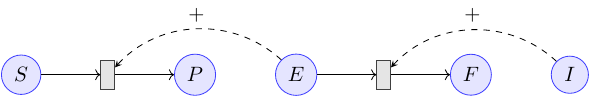}
    \caption{A qualitative Petri net representation adapted from Aduddell et al.~\cite{aduddell2024compositional}. Places (circles, e.g., S, P) represent molecular species, and transitions (boxes) represent biochemical reactions. Dashed arrows denote regulatory influences, where the ``+'' symbol signifies a positive influence or catalysis.}
    \label{fig:adu_petri_example}
\end{figure}

\begin{figure}[h!]
    \centering
    \includegraphics[width=0.9\textwidth]{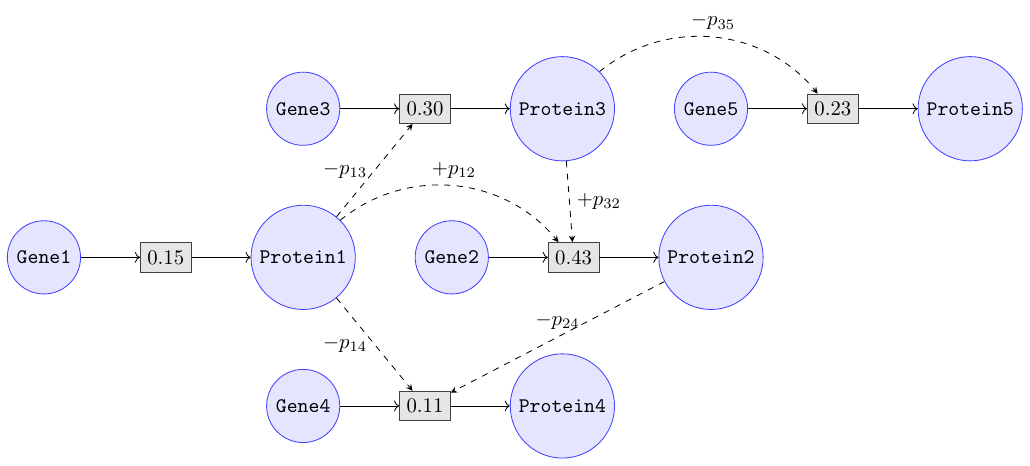}
    \caption{Our mechanistic BTPN layer. The model is structurally identical to a TPN, but each quantitative label (e.g., the base rate 0.15 or the parameters of $-p_{13}$) is now interpreted as a sample from an underlying probability distribution learned from data.}
    \label{fig:tpn_example}
\end{figure}

\begin{figure}[h!]
    \centering
    \includegraphics[width=1.0\textwidth]{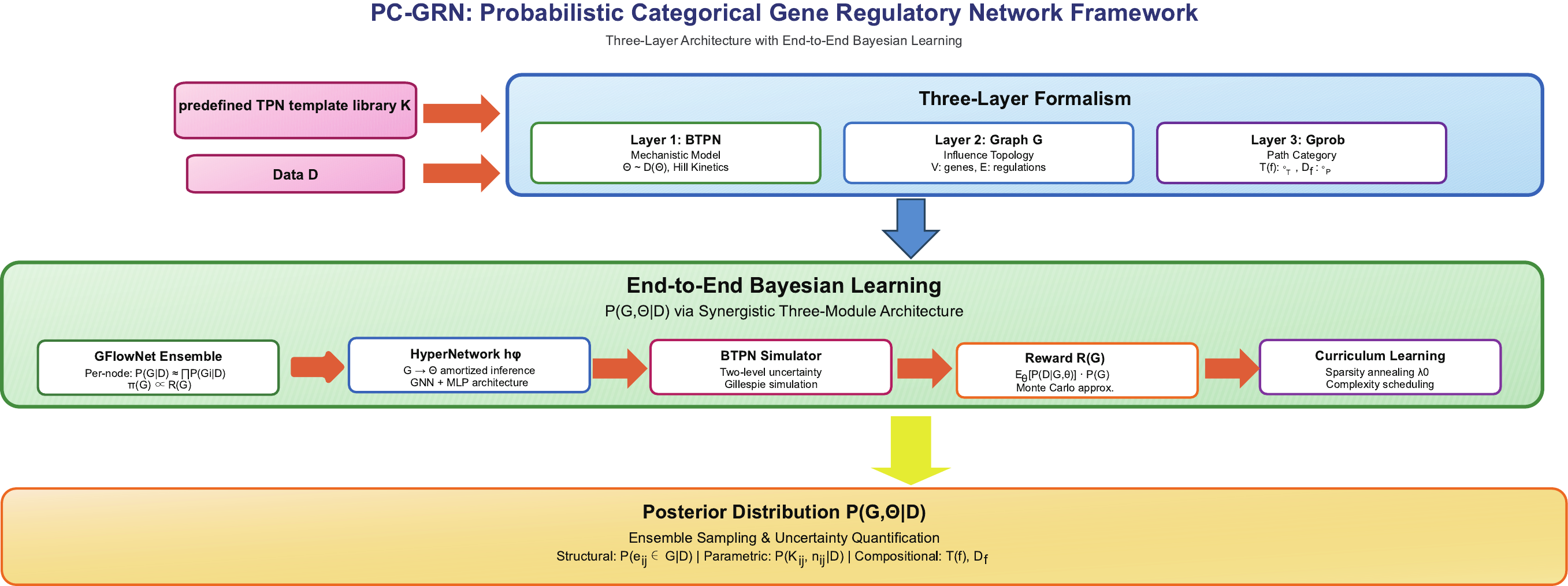}
    \caption{The overall architecture of the PC-GRN framework, illustrating the synergistic integration of its three core components. The three-layer formalism(top) defines the mathematical objects, from a mechanistic BTPN to a compositional path category ($G_{\mathrm{prob}}$). The end-to-end Bayesian learning engine (middle) is designed to learn a full posterior distribution over these objects ($P(G, \Theta | D)$) directly from data. It combines a GFlowNet for topology sampling, a HyperNetwork for parameter inference, and a BTPN simulator for evaluation to compute a reward that guides learning. The final output is the characterization of this learned posterior distribution (bottom), which enables a comprehensive, uncertainty-aware analysis of the GRN.}
    \label{fig:framework}
\end{figure}

A key aspect of our framework is how it utilizes basic components, which contrasts with purely formal approaches such as that of Aduddell et al.~\cite{aduddell2024compositional}. Their work provides a powerful, abstract syntax for describing how components like reactions (transitions) and catalysts (regulatory species) can be composed, as seen in their qualitative Petri-net representations (Figure~\ref{fig:adu_petri_example}). Our framework builds upon this formal groundwork by introducing an explicit \textbf{quantitative semantics} for the dynamics. Each transition is assigned a learnable base reaction rate ($\lambda^0$), and each regulatory interaction is defined by a specific, learnable kinetic function (such as a \textbf{Hill function}) that precisely dictates how regulator concentrations (i.e., token counts) modulate this rate (as will be illustrated in Figure~\ref{fig:tpn_example}).

Furthermore, while their framework defines the static, mathematical rules for composing predefined structures, our methodology operationalizes these components within a dynamic, \textbf{learning-based paradigm}. Our approach shifts the focus from declaratively defining composition to learning a \textbf{generative policy} for composition. The ``basic components''(fundamental biochemical reactions) are encoded in a predefined TPN template library, $\mathcal{K}$. Instead of being composed by fixed rules, this library serves as a \textbf{vocabulary} or \textbf{action space} for our generative agent. Specifically, the GFlowNet module treats network construction as a sequential decision-making process, learning a policy to sample sequences of actions that incrementally build topologies best explaining the experimental data. This generative approach is also fundamentally more flexible, as it is capable of constructing and evaluating networks with \textbf{feedback loops}, a critical feature of biological systems often disallowed in traditional acyclic graphical models.

\subsection{Formalism of the PC-GRN Framework}
\label{ssec:formalism_overview}
The PC-GRN formalism provides a rigorous, three-layered bridge from concrete biochemical processes to an abstract compositional algebra for reasoning about GRN behavior. The progression begins with a detailed BTPN, an uncertainty-aware mechanistic model where dynamic parameters are represented as probability distributions. This mechanistic model is then abstracted into a directed influence graph that captures the high-level regulatory topology. Finally, this graph generates a probabilistically enriched path category, which provides the formal language for compositional reasoning.

\subsubsection{Layer 1: The TPN as the Mechanistic Substrate}
The foundational layer of PC-GRN is BTPN, which provides an interpretable and uncertainty-aware mechanistic model. Unlike a standard TPN which is defined by a single set of parameters, a BTPN is defined by a topology and a set of \textbf{probability distributions} over its dynamic parameters. Both the topology and the hyperparameters of these distributions are inferred from data by our learning architecture.

The building blocks of BTPNs are defined as follows:
\begin{itemize}
    \item \textbf{Typed Places}: Representing gene/molecular species (e.g., \texttt{Gene1}, \texttt{Protein1}). Places hold discrete, typed \textbf{tokens}.
    
    \item \textbf{Stochastic Transitions}: Representing biochemical reactions. For a given network topology $G$, each transition $k \in G$ possesses a \textbf{learnable base rate distribution}, $D(\lambda_k^0)$, from which a specific rate can be sampled for each simulation run.
    
    \item \textbf{Probabilistic Regulatory Arcs}: Represented as dashed arrows, these signify a modulatory influence on a transition's firing rate. Each regulatory arc is associated with a set of \textbf{learnable parameter distributions}, such as $D(K)$ for interaction strength and $D(n)$ for cooperativity. The HyperNetwork's role is to predict the hyperparameters (e.g., mean and variance) that define these distributions.
\end{itemize}

The dynamics of a BTPN are inherently stochastic, reflecting both the randomness of biochemical events and the uncertainty in our knowledge of the parameters. To run a simulation, the first thing is to choose a concrete set of parameters $\theta = \{\lambda_k^0, K_{ij}, n_{ij}, ...\}$ from their respective distributions. With this sampled parameter set, the firing rate $\lambda_k$ for any transition $k$ is then calculated using the Hill function defined for Stochastic GRNs \cite{hernandez2023corrected}:

\begin{equation}
\label{eq:generalized_hill}
\lambda_k = \frac{\lambda_k^{\text{basal}} + \sum_{i \in \text{Activators}} \lambda_{A_i}^{\text{max}} \left( \frac{[\text{A}_i]}{K_{A_i}} \right)^{n_{A_i}}}{1 + \sum_{i \in \text{Activators}} \left( \frac{[\text{A}_i]}{K_{A_i}} \right)^{n_{A_i}} + \sum_{j \in \text{Inhibitors}} \left( \frac{[\text{R}_j]}{K_{R_j}} \right)^{n_{R_j}}}
\end{equation}

The components of this equation have the following biological and kinetic meanings:
\begin{itemize}
    \item \textbf{$\lambda_k$}: The final \textbf{effective firing rate} of the transition, which is the rate used by the simulator after accounting for all regulatory influences.

    \item \textbf{$\lambda_k^{\text{basal}}$}: The \textbf{basal or leaky rate} of the reaction, representing the baseline rate of expression in the absence of any dedicated activators.

    \item \textbf{The Numerator}: This term represents the \textbf{total activated production rate}. It is composed of the basal rate plus the sum of contributions from all activator species present. Each term $\lambda_{A_i}^{\text{max}}$ denotes the maximum rate contributed by a saturating concentration of activator $A_i$.

    \item \textbf{The Denominator}: This term acts as a normalization factor that accounts for the competitive binding to the promoter. It sums the relative probabilities of all possible promoter states: unbound (represented by the value 1), bound by any of the activators, and bound by any of the repressors.

    \item \textbf{$[\text{A}_i], [\text{R}_j]$}: The concentrations of activator and repressor species, respectively, which correspond to the \textbf{token counts} in their respective places in our BTPN model.

    \item \textbf{$K_{A_i}, K_{R_j}$}: The \textbf{half-effect constants} (or dissociation constants). They represent the concentration at which a regulatory effect is half-maximal and are a core measure of interaction strength (i.e., binding affinity).

    \item \textbf{$n_{A_i}, n_{R_j}$}: The \textbf{Hill coefficients}, which describe the cooperativity of the interaction. A value of $n>1$ indicates a sigmoidal, switch-like response, while $n=1$ represents a non-cooperative, hyperbolic response.
\end{itemize}

In the context of the PC-GRN framework, all parameters in this equation ($\lambda^{\text{basal}}$, $\lambda^{\text{max}}$, $K$, and $n$) are the learnable quantities. The HyperNetwork module is specifically designed to infer the posterior distributions over these parameters for any given network topology discovered by the GFlowNet, thereby defining a complete, simulatable, and biologically meaningful dynamic model.

To illustrate the application of this formula, let us consider two multi-input transitions from Figure~\ref{fig:tpn_example} as examples:

\begin{itemize}
    \item \textbf{Generation of Protein2}: This process has a base rate of 0.43 and is simultaneously activated by \texttt{Protein1} and \texttt{Protein3}. Its firing rate, $\lambda_{\text{P2}}$, can be expressed as:
    \begin{equation}
        \lambda_{\text{P2}} = 0.43 \cdot \frac{1 + \left( \frac{[\text{Protein1}]}{K_{12}} \right)^{n_{12}} + \left( \frac{[\text{Protein3}]}{K_{32}} \right)^{n_{32}}}{1 + \left( \frac{[\text{Protein1}]}{K_{12}} \right)^{n_{12}} + \left( \frac{[\text{Protein3}]}{K_{32}} \right)^{n_{32}}}
    \end{equation}

    \item \textbf{Generation of Protein4}: This process has a base rate of 0.11 and is simultaneously inhibited by \texttt{Protein1} and \texttt{Protein2}. Its firing rate, $\lambda_{\text{P4}}$, can be expressed as:
    \begin{equation}
        \lambda_{\text{P4}} = 0.11 \cdot \frac{1}{1 + \left( \frac{[\text{Protein1}]}{K_{14}} \right)^{n_{14}} + \left( \frac{[\text{Protein2}]}{K_{24}} \right)^{n_{24}}}
    \end{equation}
\end{itemize}

In this paradigm, the task of our learning architecture (GFlowNet and HyperNetwork) is to first discover the existence of these regulatory connections (e.g., \texttt{-p\textsubscript{14}}, \texttt{-p\textsubscript{24}}) and then to infer the posterior distributions for their corresponding kinetic parameters ($K_{14}, n_{14}, K_{24}, n_{24}$). These examples highlight the expressive power of our formalism, which can naturally represent the complex, multi-input regulatory logic that governs synergistic gene expression.

By running many such simulations, each with a fresh sample of parameters $\theta$, we can generate an ensemble of trajectories that fully captures the posterior predictive distribution of the system's behavior. The uncertainty represented by the parameter distributions within the BTPN is now the direct source for the uncertainty $D_f$ in the categorical layer's morphisms.

\subsubsection{Layer 2: The Abstracted Directed Influence Graph}
\begin{figure}[h!]
    \centering
    \includegraphics[width=0.5\textwidth]{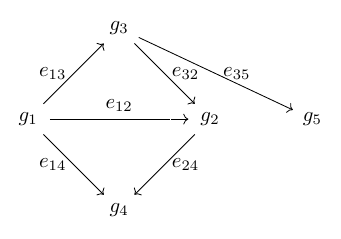}
    \caption{The abstracted directed influence graph $\mathcal{G}$, derived from the TPN in Figure~\ref{fig:tpn_example}. This graph captures the high-level regulatory topology.}
    \label{fig:graph_example}
\end{figure}

To bridge the detailed mechanistic model with the abstract compositional algebra, the BTPN is first simplified into a directed influence graph, $\mathcal{G} = (V, E)$ (Figure~\ref{fig:graph_example}). This graph discards the kinetic details and represents only the essential ``who regulates whom'' topology of the network.

\begin{itemize}
    \item \textbf{Vertices ($V$)}: The vertices $g_i \in V$ correspond to the core genes of the TPN.
    \item \textbf{Edges ($E$)}: A directed edge $e_{ij}: g_i \to g_j$ exists in $E$ if a product of gene $i$ directly modulates a transition associated with the expression of gene $j$ in the TPN.
\end{itemize}

\subsubsection{Layer 3: The Probabilistically Enriched Path Category $G_{\mathrm{prob}}$}
Finally, we define the category $G_{\mathrm{prob}}$ as the \textbf{probabilistically enriched path category} generated from the influence graph $\mathcal{G}$.

\begin{itemize}
    \item \textbf{Objects}: The objects of $G_{\mathrm{prob}}$ are the vertices of $\mathcal{G}$.    
    \item \textbf{Morphisms}: A morphism $f: g_i \to g_j$ in $G_{\mathrm{prob}}$ corresponds to a \textbf{path} in $\mathcal{G}$.    
    \item \textbf{Composition}: Composition of morphisms is the \textbf{concatenation of their corresponding paths}
    \item \textbf{Probabilistic Enrichment}: Each morphism is enriched with information derived from the underlying TPN:
    \begin{itemize}
        \item  A \textbf{type} $T(f)$ is assigned to each path. The type of a morphism corresponding to a single edge is its sign in the TPN, either positive ($+$) for \textbf{activation} or negative ($-$) for \textbf{inhibition}. The type of a longer path is determined by composing the signs of its constituent edges via a logical rule $\circ_T$. For example, the disinhibition rule is expressed as $(-) \circ_T (-) = (+)$.
        \item A \textbf{probability distribution} $D_f$ is assigned to each path. The distribution for a single-edge path is derived from the full posterior over the kinetic parameters (e.g., $K, n$) of the corresponding regulatory arc in the TPN. The distribution for a longer path is computed by propagating uncertainty via a rule $\circ_P$(see the discussion in Section~\ref{ssec:future_work}).
    \end{itemize}
\end{itemize}

\subsection{The End-to-End Bayesian Learning Architecture}
\label{ssec:learning_architecture}

To infer the posterior distribution over BTPNs, \( P(G, \Theta \mid D) \)(the joint posterior probability of a specific network topology \( G \) and its associated parameter distributions \( \Theta \) given the observed experimental data \( D \)), we propose an integrated, end-to-end Bayesian learning framework. This architecture simultaneously discovers the network structure \( G \) and learns the probability distributions over its kinetic parameters \( \Theta \), enabling a unified treatment of both structural and parametric uncertainty.

Our architecture comprises three core modules operating in a synergistic feedback loop:

\begin{enumerate}
    \item \textbf{GFlowNet as a Structure Sampler:} GFlowNet serves as the structure discovery engine. It learns a stochastic policy to sample candidate network topologies \( G \) from the vast combinatorial space of possible GRNs, with sampling probabilities proportional to the posterior probability of each topology.

    \item \textbf{HyperNetwork as a Distribution Predictor:} The HyperNetwork functions as a dynamic parameterizer. Given any topology \( G \), it instantly predicts the \textbf{hyperparameters} (e.g., mean and variance) that define the probability distributions over all kinetic parameters \( \Theta = \{D(\lambda_k^0), D(K_{ij}), D(n_{ij}), \ldots\} \). Specifically, \( D(\lambda_k^0) \) denotes the base reaction rates, \( D(K_{ij}) \) represents interaction strengths, and \( D(n_{ij}) \) captures cooperativity—collectively quantifying the full parametric uncertainty of the model. This approach eliminates the need for a costly, nested optimization loop for each proposed topology.

    \item \textbf{BTPN Simulator as an Evaluator:} The BTPN Simulator acts as the physics engine. It takes a candidate topology \( G \) and its associated parameter distributions \( \Theta \), runs stochastic simulations by sampling parameters from these distributions, and computes a scalar reward by comparing the simulation output to the experimental data \( D \).
\end{enumerate}

We conceptualize BTPN topology construction as a generative process. GFlowNet incrementally builds a network \( G \) by selecting actions from a predefined library of possible interactions \( \mathcal{K} \). This process is guided by a \textbf{reward function} \( R(G) \), which is proportional to the posterior probability \( P(G \mid D) \). This reward function allows the incorporation of prior knowledge(e.g., a preference for sparser networks), thereby steering the search toward biologically plausible GRNs. The mathematical formulation of this reward is presented in Section~\ref{sec:learning_architecture}.

To summarize, the three modules operate in a closed, synergistic loop:

\begin{enumerate}
    \item GFlowNet proposes a candidate network topology \( G \).
    \item The HyperNetwork instantly predicts the distributions for its kinetic parameters \( \Theta \).
    \item The BTPN Simulator evaluates how well the resulting stochastic BTPN \( (G, \Theta) \) explains the observed data, producing a scalar reward.
\end{enumerate}

This reward signal is then used to update both the GFlowNet's policy and the HyperNetwork's weights, allowing the entire architecture to jointly converge toward the posterior distribution. The specific algorithms driving this joint optimization are detailed in Section~4. This process can be viewed as learning a generative model for the posterior \( P(G, \Theta \mid D) \), which is subsequently approximated via \textbf{Monte Carlo sampling} of high-reward \( (G, \Theta) \) pairs to characterize the distribution over specific pathways or interactions.

\subsection{Mapping the BTPN Posterior to the Categorical Framework}
\label{sec:mapping_tpn_to_category}

The mapping from the learned BTPN posterior to the final category $G_{\mathrm{prob}}$ proceeds in two steps. First, the posterior distribution over detailed BTPN models is abstracted into a distribution over high-level influence graphs. Second, the path category derived from this ensemble of graphs is probabilistically enriched to yield $G_{\mathrm{prob}}$.

\textbf{Step 1: From BTPNs to Influence Graphs.}  The learning process produces a generative model of the joint posterior $P(G, \Theta \mid D)$. By marginalizing out the kinetic parameters $\Theta$, we obtain the posterior distribution over network topologies, $P(G \mid D)$. The trained GFlowNet acts as a sampler from this distribution, generating an ensemble of high-probability influence graphs $\{G_1, G_2, \ldots, G_N\}$—each corresponding to a concrete regulatory hypothesis sampled in proportion to $P(G \mid D)$. Each graph $G_k$ in the ensemble encodes a specific "who-regulates-whom" structure (as illustrated in Figure~\ref{fig:graph_example}). Because sampling is guided by posterior probability, the ensemble predominantly consists of plausible, data-consistent topologies. Importantly, this ensemble allows us to quantify structural uncertainty—for example, a regulatory edge that appears in 95\% of the sampled graphs is inferred to have higher posterior credibility than one that appears in only 30\%.

\textbf{Step 2: From Influence Graphs to a Probabilistic Category.}  The final category $G_{\mathrm{prob}}$ is constructed from the full ensemble of sampled graphs. Its structure is based on the concept of a path category, but its components reflect statistical properties aggregated across the ensemble.

\begin{itemize}
    \item \textbf{Objects}: The objects of $G_{\mathrm{prob}}$ are the gene nodes $\{g_i\}$ that appear across the sampled graphs.

    \item \textbf{Morphisms}: A morphism $f: g_i \to g_j$ in $G_{\mathrm{prob}}$ represents the \textbf{probabilistic ensemble of all paths} from $g_i$ to $g_j$ across the entire graph ensemble $\{G_k\}$—not just a single path in a single graph.

    \item \textbf{Morphism Type $T(f)$}: The type of a morphism (e.g., activation or repression) is determined by the \textbf{net effect} observed across the ensemble. For example, if an edge $e_{ij}$ is classified as activating in 90\% of high-reward sampled graphs, then the corresponding morphism will be labeled as activation.

    \item \textbf{Morphism Distribution $D_f$}: Each morphism is associated with a probability distribution $D_f$, modeled as a \textbf{mixture distribution} that aggregates uncertainty from the posterior. For an elementary morphism $f_{ij}$ corresponding to an edge $e_{ij}$, the distribution $D_{f_{ij}}$ is computed as a weighted average over the parameter distributions $\Theta_{ij}$, with weights given by the posterior probabilities of the graphs in which $e_{ij}$ appears. This captures both intra-structural (parametric) and inter-structural (topological) uncertainty.
\end{itemize}

This two-step mapping ensures that the final categorical representation, $G_{\mathrm{prob}}$, is rigorously grounded in the full posterior distribution over mechanistic models. It cohesively integrates both structural and parametric uncertainty into a formal compositional language.

\section{An End-to-End Bayesian Learning Architecture}
\label{sec:learning_architecture}

This section provides a detailed technical exposition of the end-to-end Bayesian learning architecture that was introduced conceptually in the previous section. While Section~\ref{sec:framework} outlined the roles of the three core modules: the GFlowNet, the HyperNetwork, and the BTPN Simulator, this section delves into their mechanics and mathematical foundations. We formally define the generative learning problem, specify the mathematical structure of the reward function, and detail the algorithms that enable the synergistic, joint inference of both network structure and parameters.

\subsection{Formulation of the Generative Learning Problem}
\label{ssec:generative_problem_formulation}
To leverage GFlowNets for structure discovery, we first formally define the problem space of BTPNs.

\textbf{BTPN Component Space:} Our learning process begins with a predefined template library that specifies the complete set of possible \textbf{places}, denoted $\mathcal{P}$, and candidate \textbf{transitions}, denoted $\mathcal{K}$. The set $\mathcal{P}$ represents typed molecular species and cellular states in the model, encompassing not only fundamental entities such as genes, mRNAs, and proteins (in their various functional states, e.g., active or inactive), but also higher-order constructs like protein complexes, regulatory small molecules, and abstract cellular conditions. The set $\mathcal{K}$ defines all plausible biochemical reactions or interactions among these entities. A specific GRN topology $G$ is formally defined as a subset of possible transitions, i.e., $G \subseteq \mathcal{K}$. Consequently, the search space for candidate network structures corresponds to the power set of $\mathcal{K}$, that is, $2^{\mathcal{K}}$.

We then construct network topologies through a sequence of actions, framing the generation of a topology $G$ as a sequential decision-making process. To illustrate, consider the construction of the network shown in Figure~\ref{fig:tpn_example}. The process begins from an initial state $s_0$, which includes the ten gene and protein places but contains no transitions or regulatory arcs. At each step $t$, the GFlowNet selects an action $a_t$ from a predefined set of allowable interactions to incrementally build the network. A possible sequence of actions might be:

\begin{itemize}
    \item $a_1$: Add the basal expression transition \texttt{Gene1} $\to$ \texttt{Protein1}.
    \item $a_2$: Add the expression transition for \texttt{Protein4}.
    \item $a_3$: Add the inhibitory regulatory arc \texttt{-p\textsubscript{14}}, linking \texttt{Protein1} to the \texttt{Protein4} transition.
\end{itemize}

Each action $a_t$ leads to an updated state, $s_t = s_{t-1} \cup \{a_t\}$. A complete action trajectory, denoted $\tau = (s_0, a_1, s_1, \dots, s_n = G)$, thus uniquely defines the resulting BTPN topology $G$.

Next, we leverage the reward function as an approximation of the marginal likelihood. The ultimate goal of our Bayesian framework is to sample topologies $G$ from the true posterior distribution, $P(G \mid D)$, which is proportional to the marginal likelihood $P(D \mid G)$ multiplied by the structural prior $P(G)$:
\[
P(G \mid D) \propto P(D \mid G) \cdot P(G)
\]
The marginal likelihood $P(D \mid G)$ provides a principled scoring function for network structures, as it quantifies the likelihood of observing the data under all possible parameterizations of $G$, weighted by their prior probabilities:
\[
P(D \mid G) = \int P(D \mid G, \Theta) \, P(\Theta \mid G) \, d\Theta
\]
However, this integral is high-dimensional and analytically intractable. To overcome this challenge, we employ a two-part strategy to define a tractable reward function $R(G)$ that closely approximates the true marginal likelihood.

First, we use a HyperNetwork to perform amortized inference as discussed in \cite{bitzer2023amortized}. Rather than integrating over the entire space of parameter distributions, the HyperNetwork $h_\phi$ learns a direct mapping from a given structure $G$ to a plausible parameter distribution $P(\Theta \mid G, \phi)$, thereby significantly reducing computational complexity.

Second, we approximate the expectation over this distribution using Monte Carlo sampling. The resulting reward function for a given structure $G$ is defined as the expected data likelihood, averaged over samples from the HyperNetwork's predicted distribution, and scaled by the structural prior:
\[
R(G) \approx \left( \mathbb{E}_{\theta \sim P(\Theta \mid G, \phi)} \left[ P(D \mid G, \theta) \right] \right) \cdot P(G)
\]
In practice, the expectation $\mathbb{E}[\cdot]$ is approximated by drawing a finite number of parameter vectors $\{\theta_1, \dots, \theta_m\}$ from $P(\Theta \mid G, \phi)$ and averaging their corresponding likelihoods. The prior $P(G)$ encodes biological assumptions, such as an $L^0$ sparsity prior,
\[
P(G) \propto \exp(-\lambda_0 \|G\|_0),
\]
which discourages overly complex networks. This reward function is both computationally tractable and robust to parametric uncertainty.

\subsection{Mechanics of Three Core Modules}
\label{ssec:stage2_inference_engine}
This section details the mechanics of the three core modules and, crucially, how they interoperate as a joint inference engine to learn the posterior distribution $P(G, \Theta \mid D)$.

A central challenge in GRN inference is the combinatorial explosion of the search space: the number of possible network topologies for $d$ genes scales as $2^{d^2}$, rendering brute-force search computationally infeasible. To address this, our framework adopts a \textbf{per-node factorization} strategy. This divide-and-conquer approach decomposes the global problem of inferring the entire network into $d$ localized subproblems—one for each gene.

Rather than employing a single, monolithic GFlowNet, we utilize an ensemble of $d$ independent GFlowNets, $\{GFN_1, \dots, GFN_d\}$, where each $GFN_i$ is responsible for learning a localized policy to construct the set of \textit{incoming} regulatory transitions for its corresponding gene $g_i$. This design is grounded in the biological principle of modularity: the regulatory logic of a gene is largely determined by its promoter-level interactions. The full network posterior is then approximated by the product of local posteriors:
\[
P(G \mid D) \approx \prod_{i=1}^{d} P(G_i \mid D)
\]
This factorization—commonly assumed in classical GRN inference algorithms—dramatically reduces the complexity and action space for each generative agent, thereby improving scalability to large systems.

Next, the HyperNetwork $h_\phi(G) \rightarrow \Theta$ functions as the core engine for parameter inference. It provides a powerful means of conducting amortized inference over kinetic parameters conditioned on arbitrary network topologies. Rather than performing computationally expensive and iterative optimization for each sampled structure, the HyperNetwork learns a single, highly expressive non-linear function that directly maps a given graph $G$ to a distribution over its associated parameters $\Theta$.

The parameter inference pipeline consists of the following components:

\begin{enumerate}
    \item \textbf{Graph Encoding:} Since network topologies ($G$) are variable-sized graphs, the HyperNetwork first employs a Graph Neural Network to encode the structural information. The GNN processes the topology of $G$ and outputs a fixed-dimensional vector representation, known as a graph embedding, which compactly summarizes the key topological features.

    \item \textbf{Parameter Prediction:} This graph embedding is then passed through a Multi-Layer Perceptron. The final layer of the MLP is designed with multiple output heads, each dedicated to predicting the hyperparameters (e.g., mean $\mu$ and variance $\sigma^2$) of a specific kinetic parameter distribution, such as $D(K_{ij})$, in the set $\Theta$.

    \item \textbf{Differentiable Gradient Flow:} This formulation allows the HyperNetwork to generate full probability distributions on-the-fly for any given graph, which is essential for computing the reward signal during training. Furthermore, the entire inference process is fully differentiable. By leveraging techniques such as the reparameterization trick during the simulation step, gradients originating from the reward signal can be propagated end-to-end through both the GNN encoder and the MLP updating the parameters $\phi$.
\end{enumerate}

Finally, the BTPN Simulator provides the mechanistic grounding for the entire framework. It evaluates the plausibility of a proposed GRN model by executing an ensemble of stochastic simulations. Given a complete model $(G, \Theta)$ and an initial condition, the simulator generates synthetic data $\hat{D}$ to compare against the observed data $D$ for reward computation.

This simulation process captures two distinct, nested forms of uncertainty:

\begin{itemize}
    \item \textbf{Level 1: Epistemic Uncertainty (Uncertainty in Knowledge).} Before each simulation run, the simulator samples a concrete parameter vector $\theta_k = \{\lambda^0, K, n, \dots\}$ from the predicted distribution $\Theta$. This accounts for our lack of knowledge about the true kinetic parameters. Each sampled $\theta_k$ represents a plausible realization of biological reality, and multiple samples allow us to explore this uncertainty space.

    \item \textbf{Level 2: Aleatoric Uncertainty (Intrinsic Stochasticity).} Given a fixed parameter vector $\theta_k$, the simulator performs a stochastic simulation(e.g., via the Gillespie algorithm) to model the intrinsic randomness in reaction timings. Even with identical parameters, different runs yield different molecular trajectories.
\end{itemize}

By repeatedly executing this two-level ``sample-then-simulate'' process, the simulator generates an ensemble of trajectories $\hat{D}$ that capture both epistemic and aleatoric uncertainties. The primary reward signal is then derived by assessing the fidelity of this ensemble to the experimental data $D$, thereby providing a robust, uncertainty-aware metric that guides both the GFlowNet and HyperNetwork toward more biologically plausible BTPN models.

\begin{figure}[h!]
    \centering
    \includegraphics[width=0.8\textwidth]{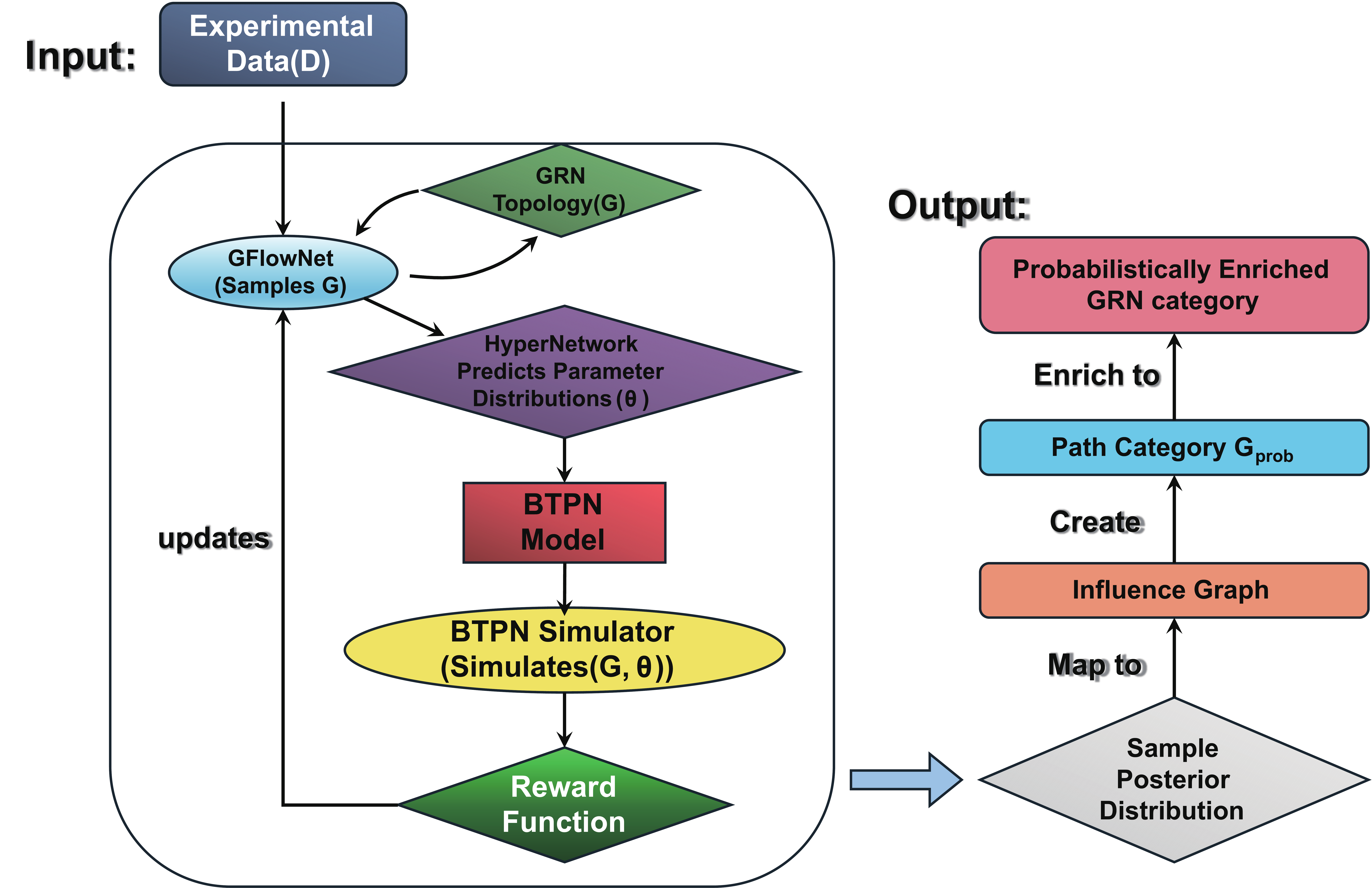}
    \caption{The end-to-end workflow of the PC-GRN framework.}
    \label{fig:workflow}
\end{figure}

\subsection{The Training Algorithm and Refinement}
\label{ssec:training_algorithm}

The proposed framework is trained end-to-end via a joint optimization procedure, which is summarized in Figure~\ref{fig:workflow} and detailed in Algorithm~\ref{alg:training_loop}. As illustrated in the workflow, the Bayesian Learning Loop is guided by experimental data ($D$). A GFlowNet samples a network topology ($G$), and a HyperNetwork predicts its parameter distributions ($\Theta$), together defining a BTPN model. The BTPN is stochastically simulated, and a reward function compares the output to $D$ to update the learning policy. The final output is derived from a systematic, layered abstraction of the learned posterior distribution, yielding a probabilistically enriched GRN category.

This training procedure operationalizes the core ideas of our framework. Each epoch of the training loop, as shown in Algorithm~\ref{alg:training_loop}, executes a sequence of key steps. First, the per-node GFlowNet ensemble samples a complete network topology, $G$. Subsequently, the HyperNetwork predicts the hyperparameters for the kinetic parameter distributions, $\Theta$, for this structure. With the BTPN model $(G, \Theta)$ specified, the simulator performs a two-level uncertainty simulation by repeatedly sampling a parameter vector $\theta_k$ from $\Theta$ (for epistemic uncertainty) and running a stochastic simulation with it (for aleatoric uncertainty). A reward $R(G)$ is then computed by approximating the marginal likelihood from the simulation results. This reward signal is used to update both the GFlowNet and the HyperNetwork. Finally, to further enhance stability, this entire process is managed by a \textbf{curriculum learning strategy} that gradually increases task difficulty during training.

\begin{algorithm}[H]
\SetAlgoLined
\caption{PC-GRN End-to-End Training Loop}
\label{alg:training_loop}
\KwIn{Data $D$, Template $\mathcal{K}$, Hyperparameters}
\KwOut{Trained GFlowNet Ensemble $\{GFN_i\}$, Trained HyperNetwork $h_\phi$}
\BlankLine
Initialize parameters for all $GFN_i$ and $h_\phi$\;
Initialize curriculum schedule: $\lambda_0(epoch) = \lambda_{0,initial} \times \exp(-decay\_rate \times epoch)$\;
\For{epoch = 1 to N}{
    \tcp{1. Sample Topology G using per-node GFN ensemble}
    Initialize empty graph $G$\;
    \For{i = 1 to d}{
        $G_i \leftarrow$ Sample incoming regulatory arcs for gene $g_i$ from $GFN_i$\;
        $G \leftarrow G \cup G_i$\;
    }
    \BlankLine
    \tcp{2. Predict Parameter Distributions}
    $\Theta \leftarrow h_\phi(G)$ \tcp*{Predict distributions hyperparameters}
    \BlankLine
    \tcp{3. Two-level uncertainty simulation}
    Initialize likelihood list $L = []$\;
    \For{k = 1 to m}{
        $\theta_k \leftarrow \text{Sample}(\Theta)$ \tcp*{Epistemic uncertainty}
        $\hat{D}_k \leftarrow \text{SimulateBTPN}(G, \theta_k)$ \tcp*{Aleatoric uncertainty}
        $L_k \leftarrow P(D|G, \theta_k)$\;
        Append $L_k$ to $L$\;
    }
    \BlankLine
    \tcp{4. Reward computation with curriculum learning}
    $P(D|G) \approx \frac{1}{m}\sum_{k=1}^{m} L_k$ \tcp*{Monte Carlo approximation}
    $R(G) \leftarrow P(D|G) \cdot \exp(-\lambda_0(epoch) ||G||_0)$\;
    \BlankLine
    \tcp{5. Update Models}
    Update $\{GFN_i\}$ using reward $R(G)$ and GFlowNet loss\;
    Update $h_\phi$ by backpropagating gradient from mean of likelihoods in $L$\;
    \BlankLine
    \tcp{6. Categorical framework mapping (periodic)}
    \If{epoch mod 50 == 0}{
        Construct $G_{prob}$ from high-reward topologies\;
    }
}
\end{algorithm}

As afore-mentioned, to enhance training stability and guide the discovery process toward biologically plausible solutions, our framework incorporates the curriculum learning schedule. In essence, early training epochs apply a large sparsity penalty $\lambda_0$ to encourage the identification of a coarse-grained network backbone comprising only the most essential regulatory interactions. As training progresses, this penalty is gradually annealed, enabling the GFlowNet and HyperNetwork to refine the network structure by incorporating more complex, context-dependent regulatory relationships.

Curriculum learning serves as a high-level training strategy that organizes the learning process from simple to complex tasks. It is not an intrinsic component of the GFlowNet itself, but rather an external scheduling mechanism embedded within the main optimization loop (Algorithm~\ref{alg:training_loop}) that dynamically adapts the learning environment over time. Specifically, our framework supports multiple, complementary curriculum strategies:

\begin{itemize}
    \item \textbf{Sparsity-First Curriculum:} The sparsity penalty $\lambda_0$ in the reward function is annealed across training epochs.
    \begin{enumerate}
        \item \textit{Phase 1 (Backbone Discovery):} A high initial value of $\lambda_0$ enforces strong sparsity, encouraging the GFlowNet to focus on discovering the most robust and essential regulatory edges.
        \item \textit{Phase 2 (Structure Refinement):} Gradual reduction of $\lambda_0$ permits the exploration of more elaborate regulatory motifs, refining the network structure atop the identified backbone.
    \end{enumerate}

    \item \textbf{Model Complexity Curriculum:} Early training restricts the model to a simplified parameter space, such as limiting the HyperNetwork to produce only non-cooperative interactions (e.g., enforcing Hill coefficients $n=1$). Once convergence is achieved on this linear regime, the constraint is relaxed to allow learning of non-linear dynamics (e.g., $n > 1$), supporting richer regulatory behaviors.
\end{itemize}

This multi-stage curriculum learning strategy improves both the stability and sample efficiency of training. By prioritizing the discovery of simpler, biologically plausible network structures in the early stages, it reduces the likelihood of overfitting to spurious patterns and provides a more stable foundation for learning complex gene regulatory mechanisms.

\subsection{Posterior Characterization and Interpretation}
\label{ssec:posterior_characterization}

The trained GFlowNet ensemble and HyperNetwork do not yield a single GRN, but instead enable sampling from the full posterior distribution $P(G, \Theta \mid D)$. This distributional perspective allows for a richer characterization of both network structure and dynamics, enabling an uncertainty-aware analysis rather than relying solely on point estimates. To characterize this posterior, we draw a large ensemble of high-reward $(G, \Theta)$ samples and compute relevant summary statistics.

For structural uncertainty, we analyze the set of sampled graphs $\{G_k\}$ to estimate the confidence associated with each potential regulatory edge. The posterior probability of a specific edge $e_{ij}$, representing a regulatory interaction from gene $i$ to gene $j$, is approximated by its empirical frequency across the sampled ensemble:
\[
P(e_{ij} \in G \mid D) \approx \frac{1}{N} \sum_{k=1}^{N} \mathbb{I}(e_{ij} \in G_k)
\]
where $\mathbb{I}(\cdot)$ denotes the indicator function. This enables the construction of a probabilistic adjacency matrix, offering a concise and interpretable summary of the most credible interactions, while distinguishing them from edges with weaker statistical support.

In addition to structural insights, we can also assess parameter uncertainty. For any high-confidence edge $e_{ij}$, we collect all associated kinetic parameter distributions $\Theta_{ij}$, as predicted by the HyperNetwork over the subset of sampled graphs containing that edge. This allows us to construct marginalized posterior distributions over quantities such as interaction strength ($K_{ij}$) and cooperativity ($n_{ij}$). Visualizing these distributions using histograms or kernel density estimates provides mechanistic insights into the regulatory dynamics. For instance, a posterior sharply peaked at $n_{ij} = 1$ suggests strong evidence for non-cooperative regulation, which may indicate a linear transcriptional response.

\subsection{Illustrative Case Study: Inferring Multi-Input Regulation}
\label{sec:case_study}
To concretely demonstrate the proposed PC-GRN architecture, we present a case study focused on inferring the structure and dynamics of a five-gene regulatory network, as illustrated in Figure~\ref{fig:tpn_example}. A key feature of this network is its complex multi-input regulatory logic. For example, the expression of \texttt{Protein2} is co-activated by both \texttt{Protein1} and \texttt{Protein3}, while \texttt{Protein4} is co-repressed by \texttt{Protein1} and \texttt{Protein2}. This case study aims to show that PC-GRN can autonomously discover such specific and nontrivial regulatory structures solely from simulated time-series gene expression data.

We begin by defining a BTPN template library that specifies the set of possible \textit{places} (e.g., \texttt{Gene1}–\texttt{Gene5}, \texttt{Protein1}–\texttt{Protein5}) and a vocabulary of potential interactions (e.g., basal expression, activation, repression). The learning objective is to infer a posterior distribution over BTPN models, $(G, \Theta)$, that accurately reproduces synthetic expression trajectories $D$ generated by stochastically simulating the ground-truth network shown in Figure~\ref{fig:tpn_example}.

The learning process is driven by a per-node GFlowNet ensemble. Consider the GFlowNet responsible for gene $g_2$ (\texttt{GFN\_2}), whose task is to infer the correct set of incoming regulatory arcs for the transition producing \texttt{Protein2}. It explores multiple hypotheses, such as:
\begin{itemize}
    \item Hypothesis 1: \texttt{Protein2} is activated only by \texttt{Protein1}.
    \item Hypothesis 2: \texttt{Protein2} is activated only by \texttt{Protein3}.
    \item Hypothesis 3 (ground truth): \texttt{Protein2} is co-activated by both \texttt{Protein1} and \texttt{Protein3}.
\end{itemize}

For each sampled topology, the HyperNetwork provides parameter distributions $\Theta$, which are used by the BTPN simulator to generate simulated data $\hat{D}$. Hypotheses that fail to include both activating regulators (e.g., 1 and 2) cannot reproduce the dynamics observed in $D$ and therefore receive low rewards. These low-reward samples update the GFlowNet policy to down-weight incorrect topologies.

In contrast, when \texttt{GFN\_2} samples the correct two-activator topology, the HyperNetwork provides appropriate parameter distributions, enabling the simulator to produce $\hat{D}$ closely matching $D$, resulting in high rewards. These rewards reinforce the GFlowNet policy to favor the correct structure, while the gradients from data-fitting errors update the HyperNetwork to refine its parameter predictions. A similar mechanism occurs with \texttt{GFN\_4} in discovering the co-repressive logic of \texttt{Protein4}.

As a result, the learning system converges to a posterior distribution $P(G, \Theta \mid D)$ concentrated on the true five-gene network topology. The trained GFlowNet ensemble samples the structure shown in Figure~\ref{fig:graph_example} with high probability, and the learned parameter distributions $\Theta$ closely match those used in data generation.

Finally, the learned mechanistic BTPN model is mapped to the categorical framework. Regulatory interactions correspond to edges in the influence graph $\mathcal{G}$ (Figure~\ref{fig:graph_example}), enabling reasoning about compositional effects. For instance, the learned model contains two regulatory paths from $g_1$ to $g_4$:
\begin{enumerate}
    \item A direct inhibitory path: $g_1 \rightarrow g_4$, represented by a morphism of type $(-)$.
    \item An indirect path via $g_2$: $g_1 \rightarrow g_2 \rightarrow g_4$, corresponding to the composition of an activation ($f_{12}$, type $+$) followed by an inhibition ($f_{24}$, type $-$), yielding a composite morphism $(-) \circ_T (+) = (-)$.
\end{enumerate}

This illustrates how PC-GRN integrates data-driven discovery of complex, multi-input mechanisms with a formal and interpretable algebraic reasoning framework for compositional regulatory effects.

\section{Discussion}
\label{sec:discussion}
In this paper, we introduced the PC-GRN framework - a novel approach aimed at tackling longstanding challenges in GRN modeling. By synergistically integrating probabilistic reasoning, categorical structure, and process-oriented semantics, PC-GRN offers a rigorous, scalable, and interpretable modeling paradigm. This section presents a critical examination of the framework’s anticipated strengths, acknowledges its current limitations, and outlines concrete directions for future development.

\subsection{Anticipated Strengths and Advantages}
\label{ssec:strengths}
The deliberate integration of category theory with the process semantics of BTPNs in a generative learning framework offers several distinct advantages over conventional GRN modeling approaches.

\noindent \textbf{Unified Treatment of Uncertainty:} PC-GRN is fundamentally designed to handle uncertainty in a principled way. Unlike methods that yield a single point estimate of the network, our framework learns the full posterior distribution \( P(G, \Theta \mid D) \), thereby simultaneously capturing both \textbf{structural uncertainty} (i.e., the existence of regulatory connections) and \textbf{parametric uncertainty} (i.e., distributions over kinetic parameters associated with those connections).

\noindent \textbf{Scalability via Compositionality and Decomposition:} The framework addresses scalability challenges from two complementary perspectives. At the learning level, the per-node GFlowNet architecture decomposes the otherwise intractable problem of inferring a global network into a series of tractable local inference tasks. At the theoretical level, the categorical formalism offers a rigorous and modular language to compose these learned modules into large-scale networks, effectively managing complexity across multiple organizational scales.

\noindent \textbf{Improved Interpretability and Biological Plausibility:} In contrast to many “black-box” machine learning models, PC-GRN emphasizes interpretability. Its outputs are not mere correlational links but fully specified mechanistic BTPN models that can be directly simulated and analyzed. Incorporating Hill-type functions and typed components anchors the model in well-established biochemical principles, ensuring that its parameters possess clear biological meaning.

\subsection{Limitations and Open Challenges}
\label{ssec:limitations}
As a conceptual framework, the practical implementation of PC-GRN entails several significant research and engineering challenges.

\noindent \textbf{Complexity of the Integrated System:} The fusion of three advanced formalisms: Category Theory, Bayesian generative modeling (via GFlowNets and HyperNetworks), and BTPNs, yields a system whose implementation and validation demand considerable expertise.

\noindent \textbf{Challenges in the End-to-End Learning System:} The successful realization of the training loop depends on addressing several non-trivial research problems:  
\begin{itemize}
  \item \textit{Training Stability:} A well-known difficulty in GFlowNet training stems from the use of a \textbf{non-stationary reward function}; the reward landscape evolves dynamically as the HyperNetwork parameters are updated~\cite{nguyen2023causal}. Achieving stable co-evolution rather than oscillatory behavior requires meticulous balancing of learning dynamics.

  \item \textit{Differentiable Simulation:} The efficiency of the training process relies on backpropagating gradients from the data-fit loss through the simulator to the HyperNetwork. Since standard stochastic simulators are non-differentiable, implementing or approximating a \textbf{differentiable BTPN simulator} (e.g., via surrogate models or differentiable physics) remains a significant technical challenge.

  \item \textit{HyperNetwork Parameterization:} The HyperNetwork must learn a highly complex mapping from discrete graph structures to the hyperparameters of a high-dimensional distribution \(\Theta\). Designing a neural architecture with sufficient capacity to model this mapping without overfitting poses a substantial challenge.
\end{itemize}

\subsection{Potential Impact and Avenues for Future Research}
\label{ssec:future_work}
Despite these challenges, we believe the PC-GRN framework holds considerable promise to advance systems biology. Our future efforts will proceed along several complementary directions:

\begin{enumerate}
    \item \textbf{Theoretical Development:} A primary objective is the rigorous formalization of robust probability composition rules (\(\circ_P\)), which governs how uncertainty propagates along multi-step pathways. While defining a general-purpose rule is a challenging open problem, several promising avenues for its definition can be explored:

\begin{itemize}
    \item \textbf{Monte Carlo Composition:} A flexible, non-parametric approach where the composite distribution is empirically constructed by repeatedly sampling from the input distributions ($D_f, D_h$) and combining these samples.

    \item \textbf{Analytical Composition:} For specific, well-behaved families of distributions, such as the log-normal distribution, an analytical solution for composition may exist, allowing for the exact calculation of the composite distribution's parameters.

    \item \textbf{Learnable Meta-Models:} Perhaps the most powerful direction is to treat the composition rule $\circ_P$ itself as a learnable meta-model. A dedicated neural network could be trained on large-scale network data to learn a function that maps the distributions of two consecutive morphisms to the empirically observed distribution of their end-to-end effect.
\end{itemize}

Beyond this foundation, our framework enables several sophisticated theoretical extensions:
    \begin{itemize}
        \item \textbf{Unification of Epistemic and Aleatoric Uncertainty:} Currently, the BTPN model primarily captures \textit{epistemic uncertainty}, which reflects our incomplete knowledge of parameters through probability distributions. However, as demonstrated by \cite{hernandez2023corrected} \textit{aleatoric uncertainty}, the system’s intrinsic stochasticity, also modulates the effective form of regulatory functions. A significant future direction is to incorporate these \textbf{stochastic correction terms} into the BTPN simulator. By accounting for real-time fluctuations in molecular concentrations during simulation, our framework can more accurately model stochastic gene expression, thereby unifying both uncertainty types within a single coherent model.

        \item \textbf{Enhanced Compositional Structures:} To fully realize the scalability inherent in our framework, the composition of learned BTPN modules may be formalized via the theory of \textbf{double categories} or \textbf{structured cospans}. Moreover, the algebraic framework of \textbf{operads} can be employed to represent multi-input interactions with greater structural hierarchy and clarity.
    \end{itemize}

    \item \textbf{Algorithmic and Software Implementation:} A critical next step involves developing concrete algorithms for the learning system and delivering a user-friendly, open-source software toolkit to facilitate adoption within the community.

    \item \textbf{Empirical Validation:} The framework must be rigorously benchmarked against established datasets and real-world biological case studies, employing standard metrics to objectively compare its performance to existing state-of-the-art approaches.
\end{enumerate}

By pursuing these avenues, we aim to transform the conceptual potential of the PC-GRN framework into a practical and impactful tool for the systems biology community.

\section{Conclusion}
\label{sec:conclusion}

Modeling GRNs remains a central challenge, demanding a framework that can simultaneously navigate combinatorial complexity, mechanistic dynamics, and profound uncertainty. This manuscript introduced the PC-GRN framework, a novel theoretical foundation designed to meet this challenge. By synergistically integrating a categorical language for composition, BTPNs for mechanistic modeling, and a generative learning engine, PC-GRN provides a unified and principled solution for inferring GRNs from data.

The core contribution of our work is a fully integrated, end-to-end Bayesian architecture that learns a joint posterior distribution over both network structure and parameter distributions, $P(G, \Theta | D)$. We achieve this through the novel combination of a GFlowNet, which learns a policy to sample high-probability network topologies ($G$), and a HyperNetwork, which performs amortized inference to predict the corresponding kinetic parameter distributions ($\Theta$). This learned, uncertainty-aware BTPN model is then abstracted via a functorial mapping into a probabilistically enriched path category, $G_{\mathrm{prob}}$. This final representation provides a formal, interpretable algebra for reasoning about the compositional effects of regulatory pathways, grounding abstract logic in a data-driven, mechanistic foundation.

PC-GRN represents a new paradigm that shifts the focus from inferring a single, static network to learning a generative model over a rich, dynamic, and probabilistic model space. As we have discussed, significant and exciting future work lies ahead, particularly in the theoretical development of probability composition rules ($\circ_P$), the algorithmic implementation of a robust software toolkit, and the unification of epistemic and aleatoric uncertainty within the BTPN simulator. By pursuing these avenues, we aim to translate the conceptual promise of PC-GRN into a powerful and practical tool for advancing our systems-level understanding of gene regulation.

\section*{Acknowledgments}
We thank the anonymous reviewers for their valuable feedback and suggestions that helped improve the quality of this work.

\section*{Funding}
This research was supported in part by [specify funding sources].

\section*{Data Availability Statement}
The datasets and code used in this study will be made available upon publication.

\section*{Conflicts of Interest}
The authors declare no conflicts of interest.



\begin{thebibliography}{99}

\bibitem{Gene_regulatory_networks}
Eric Davidson and Michael Levin.
\newblock Gene regulatory networks.
\newblock {\em Proceedings of the National Academy of Sciences}, 102(14):4935--4935, 2005.

\bibitem{Current_approaches_to_GRNs}
T~Schlitt and A~Brazma.
\newblock Current approaches to gene regulatory network modelling.
\newblock {\em BMC Bioinformatics}, 8, 2007.

\bibitem{modeling_and_analysis_of_GRNs}
Gilles Bernot, Jean-Paul Comet, Adrien Richard, Madalena Chaves, Jean-Luc Gouzé, and Frédéric Dayan.
\newblock Modeling and analysis of gene regulatory networks.
\newblock In Frédéric Cazals and Pierre Kornprobst, editors, {\em Modeling in Computational Biology and Biomedicine: A Multidisciplinary Endeavor}, pages 47--80. Springer Berlin Heidelberg, Berlin, Heidelberg, 2013.

\bibitem{alon2019introduction}
Uri Alon.
\newblock {\em An introduction to systems biology: design principles of biological circuits}.
\newblock Chapman and Hall/CRC, 2019.

\bibitem{tyson2019modeling}
John~J Tyson, Teeraphan Laomettachit, and Pavel Kraikivski.
\newblock Modeling the dynamic behavior of biochemical regulatory networks.
\newblock {\em Journal of Theoretical Biology}, 462:514--527, 2019.

\bibitem{Huang2005}
Sui Huang, Gabriel Eichler, Yaneer Bar-Yam, and Donald~E. Ingber.
\newblock Cell fates as high-dimensional attractor states of a complex gene regulatory network.
\newblock {\em Phys. Rev. Lett.}, 94(12):128701, Apr 2005.

\bibitem{Simon2007}
Simon Rosenfeld.
\newblock Stochastic cooperativity in non-linear dynamics of genetic regulatory networks.
\newblock {\em Mathematical Biosciences}, 210(1):121--142, 2007.

\bibitem{Grant2006}
Gregory Grant.
\newblock The act domain: A small molecule binding domain and its role as a common regulatory element.
\newblock {\em The Journal of biological chemistry}, 281:33825--9, 12 2006.

\bibitem{DE}
Hidde {de Jong}, Jean-Luc Gouzé, Céline Hernandez, Michel Page, Tewfik Sari, and Johannes Geiselmann.
\newblock Qualitative simulation of genetic regulatory networks using piecewise-linear models.
\newblock {\em Bulletin of Mathematical Biology}, 66(2):301--340, 2004.

\bibitem{Bayesian_Networks}
Luxuan Qu, Zhiqiong Wang, Yueyang Huo, Yuezhou Zhou, Junchang Xin, and Wei Qian.
\newblock Distributed local bayesian network for gene regulatory network reconstruction.
\newblock In {\em 2020 6th International Conference on Big Data Computing and Communications (BIGCOM)}, pages 131--139, 2020.

\bibitem{aduddell2024compositional}
Rebekah Aduddell, James Fairbanks, Amit Kumar, Pablo~S Ocal, Evan Patterson, and Brandon~T Shapiro.
\newblock A compositional account of motifs, mechanisms, and dynamics in biochemical regulatory networks.
\newblock {\em Compositionality}, 6, 2024.

\bibitem{vijesh2013modeling}
Nedumparambathmarath Vijesh, Swarup~Kumar Chakrabarti, Janardanan Sreekumar, et~al.
\newblock Modeling of gene regulatory networks: A review.
\newblock {\em Journal of Biomedical Science and Engineering}, 6(02):223, 2013.

\bibitem{xiao2009tutorial}
Yufei Xiao.
\newblock A tutorial on analysis and simulation of boolean gene regulatory network models.
\newblock {\em Current genomics}, 10(7):511--525, 2009.

\bibitem{lahdesmaki2003learning}
Harri Lähdesm{\"a}ki, Ilya Shmulevich, and Olli Yli-Harja.
\newblock On learning gene regulatory networks under the boolean network model.
\newblock {\em Machine learning}, 52(1):147--167, 2003.

\bibitem{albert2004boolean}
Réka Albert.
\newblock Boolean modeling of genetic regulatory networks.
\newblock In {\em Complex networks}, pages 459--481. Springer, 2004.

\bibitem{de2006qualitative}
Hidde de~Jong and Delphine Ropers.
\newblock Qualitative approaches to the analysis of genetic regulatory networks.
\newblock {\em System Modeling in Cellular Biology: From Concepts to Nuts and Bolts}, pages 125--147, 2006.

\bibitem{puvsnik2022review}
Žiga Pušnik, Miha Mraz, Nikolaj Zimic, and Miha Moškon.
\newblock Review and assessment of boolean approaches for inference of gene regulatory networks.
\newblock {\em Heliyon}, 8(8), 2022.

\bibitem{shmulevich2002boolean}
Ilya Shmulevich, Edward~R Dougherty, and Wei Zhang.
\newblock From boolean to probabilistic boolean networks as models of genetic regulatory networks.
\newblock {\em Proceedings of the IEEE}, 90(11):1778--1792, 2002.

\bibitem{bornholdt2008boolean}
Stefan Bornholdt.
\newblock Boolean network models of cellular regulation: prospects and limitations.
\newblock {\em Journal of the Royal Society interface}, 5(suppl\_1):S85--S94, 2008.

\bibitem{marku2023time}
Malvina Marku and Vera Pancaldi.
\newblock From time-series transcriptomics to gene regulatory networks: A review on inference methods.
\newblock {\em PLoS computational biology}, 19(8):e1011254, 2023.

\bibitem{parmar2017time}
Kiresh Parmar.
\newblock {\em Time-delayed models of genetic regulatory networks}.
\newblock PhD thesis, University of Sussex, 2017.

\bibitem{yang2020overview}
Bin Yang and Yuehui Chen.
\newblock Overview of gene regulatory network inference based on differential equation models.
\newblock {\em Current Protein and Peptide Science}, 21(11):1054--1059, 2020.

\bibitem{karlebach2008modelling}
Guy Karlebach and Ron Shamir.
\newblock Modelling and analysis of gene regulatory networks.
\newblock {\em Nature reviews Molecular cell biology}, 9(10):770--780, 2008.

\bibitem{liu2016inference}
Fei Liu, Shao-Wu Zhang, Wei-Feng Guo, Ze-Gang Wei, and Luonan Chen.
\newblock Inference of gene regulatory network based on local bayesian networks.
\newblock {\em PLoS computational biology}, 12(8):e1005024, 2016.

\bibitem{banf2017computational}
Michael Banf and Seung~Y Rhee.
\newblock Computational inference of gene regulatory networks: approaches, limitations and opportunities.
\newblock {\em Biochimica et Biophysica Acta (BBA)-Gene Regulatory Mechanisms}, 1860(1):41--52, 2017.

\bibitem{yu2017inference}
Bin Yu, Jia-Meng Xu, Shan Li, Cheng Chen, Rui-Xin Chen, Lei Wang, Yan Zhang, and Ming-Hui Wang.
\newblock Inference of time-delayed gene regulatory networks based on dynamic bayesian network hybrid learning method.
\newblock {\em Oncotarget}, 8(46):80373, 2017.

\bibitem{zou2005new}
Min Zou and Suzanne~D Conzen.
\newblock A new dynamic bayesian network (dbn) approach for identifying gene regulatory networks from time course microarray data.
\newblock {\em Bioinformatics}, 21(1):71--79, 2005.

\bibitem{dondelinger2012dynamic}
Frank Dondelinger, Dirk Husmeier, and Sophie Lèbre.
\newblock Dynamic bayesian networks in molecular plant science: inferring gene regulatory networks from multiple gene expression time series.
\newblock {\em Euphytica}, 183(3):361--377, 2012.

\bibitem{xu2007inference}
Rui Xu, Donald Wunsch~II, and Ronald Frank.
\newblock Inference of genetic regulatory networks with recurrent neural network models using particle swarm optimization.
\newblock {\em IEEE/ACM Transactions on Computational Biology and Bioinformatics}, 4(4):681--692, 2007.

\bibitem{raza2016recurrent}
Khalid Raza and Mansaf Alam.
\newblock Recurrent neural network based hybrid model for reconstructing gene regulatory network.
\newblock {\em Computational biology and chemistry}, 64:322--334, 2016.

\bibitem{shu2021modeling}
Hantao Shu, Jingtian Zhou, Qiuyu Lian, Han Li, Dan Zhao, Jianyang Zeng, and Jianzhu Ma.
\newblock Modeling gene regulatory networks using neural network architectures.
\newblock {\em Nature Computational Science}, 1(7):491--501, 2021.

\bibitem{zhao2022hybrid}
Mengyuan Zhao, Wenying He, Jijun Tang, Quan Zou, and Fei Guo.
\newblock A hybrid deep learning framework for gene regulatory network inference from single-cell transcriptomic data.
\newblock {\em Briefings in bioinformatics}, 23(2):bbab568, 2022.

\bibitem{zhao2021comprehensive}
Mengyuan Zhao, Wenying He, Jijun Tang, Quan Zou, and Fei Guo.
\newblock A comprehensive overview and critical evaluation of gene regulatory network inference technologies.
\newblock {\em Briefings in bioinformatics}, 22(5), 2021.

\bibitem{hernandez2023corrected}
Manuel~Eduardo Hernández-García and Jorge Velázquez-Castro.
\newblock Corrected hill function in stochastic gene regulatory networks.
\newblock {\em arXiv preprint arXiv:2307.03057}, 2023.

\bibitem{bitzer2023amortized}
Matthias Bitzer, Mona Meister, and Christoph Zimmer.
\newblock Amortized inference for gaussian process hyperparameters of structured kernels.
\newblock In {\em Uncertainty in Artificial Intelligence}, pages 184--194. PMLR, 2023.

\bibitem{nguyen2023causal}
Trang Nguyen, Alexander Tong, Kanika Madan, Yoshua Bengio, and Dianbo Liu.
\newblock Causal discovery in gene regulatory networks with gflownet: towards scalability in large systems.
\newblock In {\em NeurIPS 2023 Generative AI and Biology (GenBio) Workshop}, 2023.

\end{thebibliography}
\end{document}